\documentclass[onecolumn,aps,prd,preprintnumbers,superscriptaddress,nofootinbib,amsmath,amssymb,floats,floatfix,showkeys,notitlepage,longbibliography]{revtex4-1}

\usepackage{orcidlink}
\usepackage{graphicx}
\usepackage{subfigure}
\usepackage{braket}
\usepackage[utf8]{inputenc}
\usepackage{sans}
\usepackage{changes}
\usepackage{hyperref}
\hypersetup{colorlinks=true,linkcolor=blue,urlcolor=blue,citecolor=blue}
\usepackage[toc,page]{appendix}
\usepackage[normalem]{ulem}
\usepackage{adjustbox}
\usepackage{latexsym}
\usepackage{amsfonts}
\usepackage{dcolumn}
\usepackage{bm}
\usepackage{tikz}
\usepackage{bigints}
\usepackage{comment}
\usepackage{array,tabularx,multirow,booktabs}
\usepackage[tracking=true]{microtype}
\SetTracking{}{500}
\SetTracking{encoding={*}, shape=sc}{40}
\UseRawInputEncoding 
\allowdisplaybreaks

\begin{document} 
\sloppy
\title{Schwarzschild-like Black Holes Submerged in an Exponential Density Dark Matter Profile}

\author{K. Boshkayev\orcidlink{0000-0002-1385-270X}}
\email{kuantay@mail.ru}
\affiliation{National Nanotechnology Laboratory of Open Type,  Almaty 050040, Kazakhstan.}
\affiliation{Al-Farabi Kazakh National University, Al-Farabi av. 71, 050040 Almaty, Kazakhstan}
\affiliation{Kazakh‐-British Technical University, Tole bi str.~59, Almaty, 050000, Kazakhstan}

\author{Y. Sekhmani\orcidlink{0000-0001-7448-4579}}
\email{sekhmaniyassine@gmail.com}
\affiliation{Al-Farabi Kazakh National University, Al-Farabi av. 71, 050040 Almaty, Kazakhstan}
\affiliation{Center for Theoretical Physics, Khazar University, 41 Mehseti Street, Baku, AZ1096, Azerbaijan}
\affiliation{Fesenkov Astrophysical Institute, Observatory 23, 050020 Almaty, Kazakhstan}

\author{S. Zare \orcidlink{0000-0003-0748-3386}}
\email{soroushzrg@gmail.com}
\affiliation{Al-Farabi Kazakh National University, Al-Farabi av. 71, 050040 Almaty, Kazakhstan}
\affiliation{Helsinki Institute of Physics, University of Helsinki, P.O. Box 64, FI-00014 Helsinki, Finland}


\author{H. Hassanabadi\orcidlink{0000-0001-7487-6898}}
\email{hha1349@gmail.com}
\affiliation{Physics Department, California State University, Fresno, CA 93740, USA}
\affiliation{Department of Physics, University of Hradec Kr\'{a}lov\'{e}, Rokitansk\'{e}ho 62, 500 03 Hradec Kr\'{a}lov\'{e}, Czechia}
\affiliation{Al-Farabi Kazakh National University, Al-Farabi av. 71, 050040 Almaty, Kazakhstan}
\affiliation{Khazar University, Department of Physics and Electronics, 41 Mahsati Str, AZ1096, Baku, Azerbaijan}
\author{S. Mamedov \orcidlink{0000-0002-0261-3914}}
\email{ctp@khazar.org}
\affiliation{Center for Theoretical Physics, Khazar University, 41 Mehseti Street, Baku, AZ1096, Azerbaijan}
\affiliation{Institute for Physical Problems, Baku State University, Z.Khalilov 23, Baku, AZ 1148, Azerbaijan}
\affiliation{Institute of Physics, Ministry of Science and Education, H.Javid 33, Baku, AZ 1143, Azerbaijan}

\author{Ye. ~\surname{Kurmanov}}
\email[]{kurmanov.yergali@kaznu.kz}
\affiliation{National Nanotechnology Laboratory of Open Type,  Almaty 050040, Kazakhstan.}
\affiliation{Al-Farabi Kazakh National University, Al-Farabi av. 71, 050040 Almaty, Kazakhstan}

\date{\today}

\begin{abstract}
We study a class of Schwarzschild black holes embedded in an exponential-spheroidal dark matter halo, modelled by a phenomenological density profile $\rho(r)=\rho_0 e^{-r/r_0}$. By solving the Einstein equations for a static, spherically symmetric spacetime, we obtain an analytic solution for the lapse function that reduces to the Schwarzschild spacetime in the absence of the halo and to a regular halo configuration when the central black hole mass vanishes. Indeed, the two halo parameters, $\rho_0$ and $r_0$, describe the strength and radial extent of the dark matter distribution. We analyse the curvature structure, energy conditions, shadow observables, scalar quasi-normal modes and grey-body bounds of the resulting spacetime. The Ricci scalar and the Ricci square remain finite at the origin, whilst the Kretschmann scalar retains the usual central tidal singularity in the presence of a black hole mass. The weak, null, and dominant energy conditions are satisfied, whilst the strong energy condition is violated on a finite radial interval. We also show that the halo monotonically shifts the photon sphere and the shadow radius, which allows us to derive approximate constraints from the EHT observations of M87* and Sgr A*. For scalar perturbations, the Padé-resummed WKB approximation yields stable quasinormal frequencies, while the standard WKB approximation becomes unreliable for higher overtones and strong-halo configurations. Finally, the greybody bounds indicate that the halo weakens transmission through the effective barrier, particularly at low frequencies.

\end{abstract}


\maketitle

\section{Introduction}
Black holes (BHs) are among the most remarkable predictions of Einstein's field equations in General Relativity (GR). Over the past century, GR has been subjected to an extensive range of observational and experimental tests spanning both the weak- and strong-gravity regimes, and its predictions have been confirmed with remarkable precision. In the weak-field regime, this success is exemplified by classical phenomena such as the deflection of light by massive bodies \cite{misner1973gravitation,Bartelmann:2010fz}, the anomalous perihelion advance of Mercury \cite{Park:2017zgd}, and the gravitational redshift of electromagnetic signals \cite{Hohensee:2011wt}. In addition, the Shapiro time delay is regarded as the fourth classical test of GR and has found numerous applications in astrophysics, particularly in pulsar timing and precision tests of gravity \cite{shapiro1964fourth,2016ApJ...820..139C}. Another, much less familiar prediction is the Shirokov effect. Although it has received comparatively little attention, recent studies suggest that it may have promising applications in space navigation, relativistic astrometry, and other areas of space research \cite{shirokov1973one,2025GReGr..57...91M,2026arXiv260501639I,2026arXiv260623912I}.

In the strong-field realm, compact systems, most notably binary pulsars, have provided high-precision confirmation through the measured orbital decay attributed to the emission of gravitational radiation. 

Recent technological advances have enabled far more direct and sophisticated tests of GR. The LIGO-Virgo detections of gravitational waves \cite{LIGOScientific:2016aoc} have opened a new observational channel into dynamical, strong-field events, supplying direct evidence for black-hole coalescences and allowing tight probes of spacetime geometry under extreme conditions. Similarly, the Event Horizon Telescope (EHT) delivered the first resolved image of a black hole shadow \cite{EventHorizonTelescope:2019dse}, giving unprecedented observational access to the immediate near-horizon region where gravity is strongest. Long-term monitoring of stars orbiting the Milky Way's center has produced another milestone. In particular, the star S2-tracked for decades with high-precision astrometric and spectroscopic measurements \cite{Ghez:2003rt,Nagao:2017kpw,vanPutten:2017lik,Laor:1993wf,Eisenhauer:2003he}  serves as an outstanding probe of the environment near Sgr A* \cite{Boehle:2016imz}. Its precise positions and radial velocities have been exploited to infer the mass and distance of the central compact object, to set limits on possible dark-matter spikes \cite{Lacroix:2018zmg}, to test for anomalous orbital precession predicted by modified-gravity models \cite{DellaMonica:2021xcf}, to map the mass distribution in the Galactic nucleus \cite{GRAVITY:2021xju}, to search for scalar-field configurations around Sgr A* \cite{GRAVITY:2023cjt}, and to confront scenarios involving quantum-corrected BHs \cite{Xamidov:2025oqx}.

Dark matter remains one of the most enigmatic forms of matter in contemporary physics: it does not emit, absorb, or scatter electromagnetic radiation, which makes direct detection exceptionally challenging. It is commonly modeled as effectively collisionless, coupling to baryonic matter primarily through gravity, and its existence is inferred from indirect signals such as galaxy rotation curves and gravitational lensing \cite{Wright:1999jc,Dumet-Montoya:2013goa,haghi2018rotation}. The large-scale arrangement and evolution of dark matter are most fruitfully explored with $N$-body simulations, which indicate that galaxies are embedded within extended dark halos \cite{vogelsberger2020cosmological,angulo2022large}. These halos are often well described by the Navarro-Frenk-White (NFW) form, a broken power-law profile in radius \cite{Moore:1997sg,Diemand:2004wh,Jing:1998xj}. Beyond NFW, a wide variety of alternative halo prescriptions motivated by simulations and observations \cite{
Plummer:1911zza,Buyle:2006kh,deZeeuw:1985sk,1970AJ.....75...13P,Jaffe:1983iv,Hernquist:1990be}, have been proposed to capture different radial behaviors of the density. In this context, Hong-Sheng Zhao introduced a five-parameter generalized double power-law profile \cite{Zhao:1995cp} that unifies and reproduces most commonly used halo models.

Black holes situated inside galactic dark matter halos offer a controlled setting to study the two-way interaction between nonbaryonic matter and strong field spacetimes, and to exploit gravitational observables as probes of halo microphysics. In the standard cosmological picture, dark matter is the main driver of structure formation from subgalactic to cluster scales \cite{Planck:2018vyg}. At galactic distances, independent lines of evidence-rotation curves, strong and weak lensing, and dynamical measurements demand extended mass components beyond the visible matter \cite{deBlok:2009sp,Donato:2009ab}. Simple systems such as an Einstein cluster of collisionless orbiting particles naturally produce steep, centrally enhanced dark matter profiles that can affect relativistic phenomena in galactic nuclei \cite{Maeda:2024tsg}, and a covariant vector field action has been shown to reproduce this behavior while allowing a principled discussion of equilibrium and stability \cite{Fernandes:2025lon}. Work quantifying black hole appearance and gravitational-wave signatures in the presence of generic halo models demonstrates that images and ring-down spectra shift in ways that depend on the halo profile \cite{Figueiredo:2023gas}. Explicit black hole spacetimes embedded in Dekel-Zhao halos have been constructed and analyzed for their geodesics, lensing features, and quasinormal modes \cite{Ovgun:2025bol}, while Event Horizon Telescope data have been used to place limits on black hole solutions surrounded by dark matter, even when a generalized uncertainty principle minimal length is included \cite{Ovgun:2023wmc}. Building on the Burkert-halo model, as detailed in Ref. \cite{Yang:2025pmv}, a static, spherically symmetric black hole metric has been constructed within a cored Burkert halo. It has been demonstrated that the parameters of the halo lead to measurable shifts in the quasinormal-mode spectrum. Conversely, a recent study \cite{Al-Badawi:2024asn} investigated a black hole embedded in a Dehnen-type dark matter halo and found that increasing the halo parameters, namely the characteristic density $\rho_s$ and the scale radius $r_s$, weakens the gravitational barrier, shifts both the effective potential and the ISCO to larger radii, and causes particle trajectories to evolve from escaping to bound and eventually to captured orbits, with an intermediate regime exhibiting chaotic behavior. These findings highlight the significant influence of dark matter halos on particle dynamics in the vicinity of supermassive black holes.

Following Levkov et al. \cite{Levkov:2018kau}, who showed that bosonic dark matter can form self-gravitating clumps in galactic environments, we assume here that such condensates can arise on timescales short enough to be dynamically relevant and therefore model the dark component as a relativistic fluid. To capture the halo’s influence on the Schwarzschild background, we employ a central density prescription that simultaneously reproduces the inner kinematics and the outer decline of the rotation curve, specifically, the exponential-sphere profile introduced by Sofue \cite{Sofue:2013kja}. Within the exponential-halo framework, recent studies \cite{Boshkayev:2018sbj} have constructed horizonless, compact concentrations of weakly interacting particles modeled as self-gravitating, particle-dominated fluids whose outer density matches standard dark matter profiles and reported two results of direct relevance to our work:  (i) the gravitational potential of these configurations can reproduce the observed inner rotation curve of the Milky Way; and (ii) the motion of test particles (stars) at \(r\gtrsim 10^2\ \mathrm{AU}\) is effectively indistinguishable from the Schwarzschild prediction, with measurable departures appearing only at smaller radii. These findings support our modeling of the dark component as a relativistic fluid with an exponential sphere central density and motivate searches for sub-100 $\mathrm{AU}$ astrometric or spectroscopic signatures that could discriminate horizonless compact matter distributions from true black holes.

Embedding a black hole inside a galactic halo raises two closely related issues. The first is how the halo's stress-energy alters the spacetime's redshift and shape functions in a static, spherically symmetric configuration. The second is how those geometric changes show up in observables tied to null and timelike trajectories, for instance, the quasinormal mode spectrum, thermodynamics, sparsity of Hawking radiation and possible shifts in shadow signatures. Our practical approach is to begin with a halo only geometry whose redshift function is fixed by the measured circular velocity, then solve the Einstein equations including the dark matter stress-energy while imposing a Schwarzschild boundary condition at the center \cite{Matos:2000ki,Xu:2018wow}. This procedure produces an analytic metric that (i) smoothly reduces to Schwarzschild when the halo is absent and (ii) exposes explicitly how the halo's energy-momentum renormalizes the effective potential experienced by perturbations.

The paper is organized as follows. In Section 2 we derive a static, spherically symmetric black-hole solution immersed in an exponential-spheroid dark-matter halo and examine the associated energy conditions and curvature singularity structure. Section 3 addresses the thermodynamic properties, computing the Hawking temperature and heat capacity to assess local stability. In Section 4 we study the sparsity of Hawking radiation and identify the regimes in which the system behaves approximately like a blackbody. Section 5 analyzes shadow properties with reference to M87 and Sgr A*, while Section 6 is dedicated to a detailed linear-response analysis of scalar perturbations using the third-order Padé method. Finally, Section 7 summarizes the main results and their implications.

\section{Schwarzschild-like Black Holes in Dark Matter Halo}

To obtain explicit, solvable field equations we specialize to a static, spherically symmetric spacetime and impose the standard diagonal ansatz. In the coordinates \((t,r,\theta,\varphi)\) we take
\begin{equation}\label{eq:metric_ansatz}
\mathrm{d}s^{2}=-\mathcal{A}(r)\,\mathrm{d}t^{2}+\frac{\mathrm{d}r^{2}}{\mathcal{B}(r)}+r^{2}\left(\mathrm{d}\theta^2+\sin^2\theta\,\mathrm{d}\phi^2\right),
\end{equation}
where

\begin{equation}
\mathcal{A}(r)=\mathcal{B}(r)=1-\frac{2m(r)}{r}
\end{equation}
and $m(r)$ is the mass function.

We now solve the Einstein field equations to obtain a spherically symmetric black hole metric 
embedded in the ESM DM halo. 
\footnote{Working in geometric units where $G=c=1$ (thus $\kappa=8\pi$)} we express the field equations for the pure halo spacetime as follows:
\begin{equation}\label{EE_halo}
R^{\mu}_{\nu}-\tfrac{1}{2}\delta^{\mu}_{\nu}R=\kappa T^{\mu}_{\nu},
\end{equation}
where \(T^{\mu}_{\nu}=\mathrm{diag}[-\rho,p_r,p_t,p_t]\) is the (anisotropic, in general) energy-momentum tensor of the ESM profile. For the static, spherically symmetric ansatz \eqref{eq:metric_ansatz}
the independent Einstein equations may be written in the compact form
\begin{align}
\kappa T^t_t &= \mathcal{A}\Big(\frac{1}{r^2}+\frac{1}{r}\frac{\mathcal{A}'}{\mathcal{A}}\Big)-\frac{1}{r^2},  \label{EE_tt}\\[4pt]
\kappa T^r_r &= \mathcal{A}\Big(\frac{1}{r^2}+\frac{1}{r}\frac{\mathcal{A}'}{\mathcal{A}}\Big)-\frac{1}{r^2}, \label{EE_rr}\\[4pt]
\kappa T^\theta_\theta &= \tfrac{1}{2}\mathcal{A}\Big[\frac{\mathcal{A}''\mathcal{A}}{\mathcal{A}^2}
+\frac{2}{r}\Big(\frac{\mathcal{A}'}{\mathcal{A}}\Big)
\Big]. \label{EE_thth}
\end{align}

Substituting metric functions and energy-momentum tensor components into field equation we find the following simple relation
\begin{eqnarray}
    \rho=-p_r=\frac{m'(r)}{4\pi r^2} \quad {\rm and} \quad p_t=-\frac{m''(r)}{8\pi r}
\end{eqnarray}
\noindent In this work, we adopt the ESM for the density of DM halos, a simple phenomenological profile that has proven useful in describing low-surface-brightness and dwarf galaxies. 
We analyze the appropriate ESM model by considering the following energy density \cite{Sofue:2013kja}:
\begin{equation}\label{denn}
\rho(r)=\rho_0\,e^{-r/r_0},
\end{equation}
where \(\rho_0\) and \(r_0\) denote the characteristic density and scale radius of the halo, respectively. 

Consequently, the mass profile for the system a black hole plus dark matter is defined as follows
\begin{equation}
    m(r)=\int_{M_{BH}}^{r} 4 \, \pi \, \tilde{r}^2 \rho(\tilde{r})\, d\tilde{r}=M_{BH}+M_0\Bigg[1-e^{-\frac{r}{r_0}}\left(1+\frac{r}{r_0}+\frac{r^2}{2r_0^2}\right)\Bigg], \qquad {\rm with} \qquad M_0\equiv 8\pi\rho_0 r_0^3,
\end{equation}
where $M_0$ is the total mass of the dark matter distribution (halo mass). As the radial coordinate tends to infinity $r\to\infty$, the mass functions tends to the black hole mass plus dark matter halo mass $m\to M_{BH}+M_0$, and as $r\to0$, then $m\to M_{BH}$.

Correspondingly the dressed black hole metric has a Schwarzschild-like form
\begin{equation}\label{metric}
\mathcal{A}(r)\;=\;1-\frac{2M_{BH}}{r}-\frac{2M_0}{r}\Bigg[1-e^{-\frac{r}{r_0}}\left(1+\frac{r}{r_0}+\frac{r^2}{2r_0^2}\right)\Bigg].
\end{equation} 

The metric 
\eqref{metric} therefore describes a one-parameter family (in practice two parameter: \(\rho_0,r_0\)) of Schwarzschild-like black holes ``dressed'' by the ESM halo. It satisfies the following useful consistency checks:
\begin{itemize}
  \item \(\rho_0\to0\) (no halo) yields the Schwarzschild geometry;
  \item \(M_{BH}\to0\) (no central black hole) recovers the regular pure halo spacetime; and
  \item after the normalization \(\mathcal{A}(\infty)=1\) the metric is asymptotically flat and the \(1/r\) coefficient is identified unambiguously with the ADM mass \(M\).
\end{itemize}

Graphically, Fig. \ref{ff2} illustrates the variation of the metric function concerning the spacetime variable $r$. It is evident that the lapse $\mathcal{A}(r)$ exhibits significant changes in the near-horizon region, while at large $r$, it approaches a consistent asymptotic behavior. This observation indicates that the spacetime remains asymptotically flat and resembles a Schwarzschild spacetime far from the halo. In the top-left panel, a small change in $r_0$ causes a change in $A(r)$, while a large change in $r_0$ causes a shift in the zero of the metric function. Similarly, an increase in $M_0$ causes a more significant shift in the lapse curve in the strong-field region. The $3D$ plot reveals that the extremal configuration relies on both halo parameters: the extremal mass of the black hole increases with the halo’s contribution, which means that the halo raises the threshold for the formation of the event horizon.

\subsection{CURVATURE PROPERTIES AND ENERGY CONDITIONS}
In line with our black hole solution, which supports the effect of exponential dark halos, we are led to study curvature singularities governed by the Ricci scalar $R$, the squared Ricci scalar $R_{\alpha\beta} R^{\alpha\beta}$ and the Kretschmann scalar $R_{\alpha\beta\gamma\delta} R^{\alpha\beta\gamma\delta}$. These scalars are of particular interest and are defined as follows:
\begin{align}
R
&=\frac{M_0}{r_0^4}e^{-\frac{r}{r_0}}(r-4r_0) ,\
\label{r1}
\\
R_{\alpha\beta}R^{\alpha\beta}
&=\frac{M_0^2}{2r_0^8}e^{-\frac{2r}{r_0}}(r^2-4rr_0+8r_0^2) ,\
\label{r2}
\\
R_{\alpha\beta\gamma\delta}R^{\alpha\beta\gamma\delta}
&=\frac{1}{r^{6}}\Bigg[48 (M_{0}+M_{\rm BH})^{2}-\frac{8 e^{-r/r_{0}} M_{0} (M_{0}+M_{\rm BH})
\left(r^{4}+2 r^{3} r_{0}+6 r^{2} r_{0}^{2}+12 r r_{0}^{3}+12 r_{0}^{4}\right)}{r_{0}^{4}} \nonumber\\
&+\frac{e^{-2r/r_{0}} M_{0}^{2}\left(r^{8}+8 r^{6} r_{0}^{2}+16 r^{5} r_{0}^{3}+36 r^{4} r_{0}^{4}
+64 r^{3} r_{0}^{5}+96 r^{2} r_{0}^{6}+96 r r_{0}^{7}+48 r_{0}^{8}\right)}{r_{0}^{8}}\Bigg],\
\label{r3}
\end{align}

\begin{figure*}[!htp]
      	\centering{
       \includegraphics[scale=0.9]{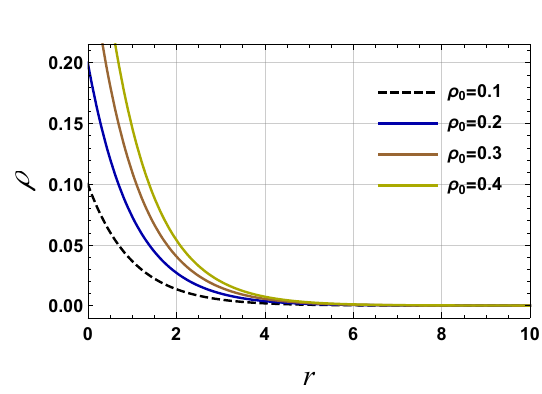} \hspace{2mm}
      	\includegraphics[scale=0.88]{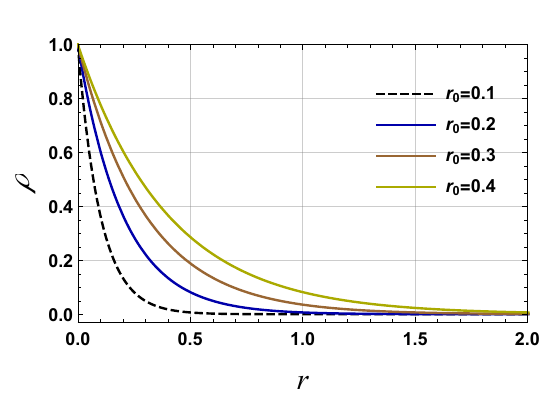} \hspace{2mm}
      	}\\
      	\centering{
       \includegraphics[scale=0.9]{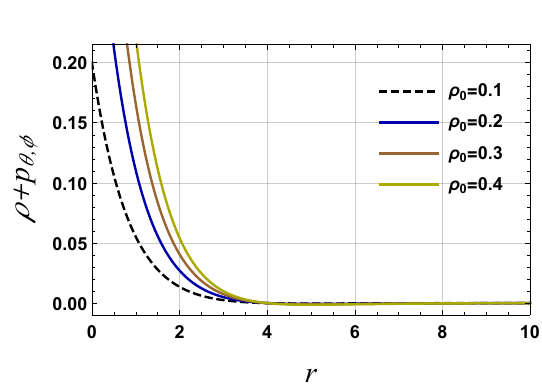} \hspace{2mm}
      	\includegraphics[scale=0.88]{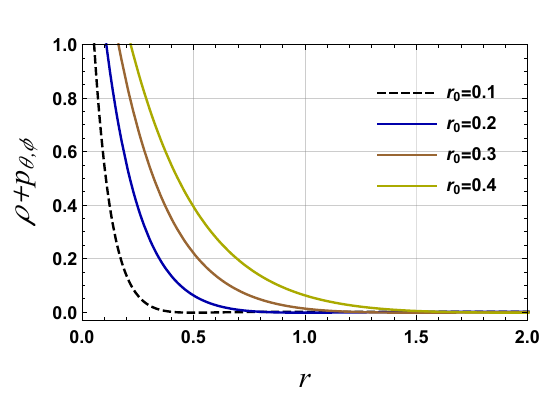} \hspace{2mm}
      }
      	\centering{
       \includegraphics[scale=0.9]{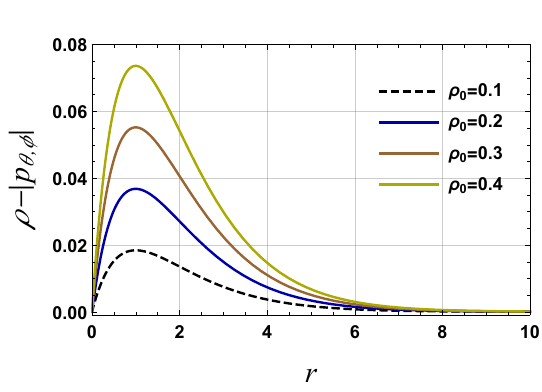} \hspace{2mm}
       \includegraphics[scale=0.88]{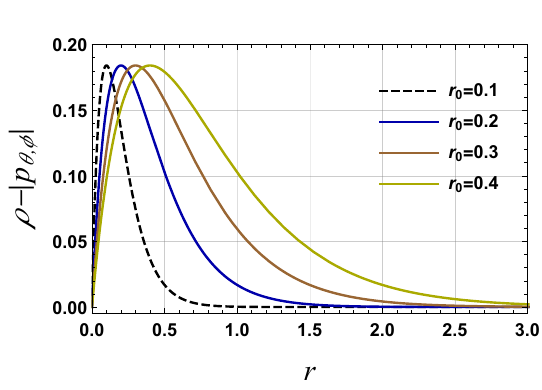} \hspace{2mm}
      	
        \hspace{2mm}
      }
      	\centering{
       \includegraphics[scale=0.91]{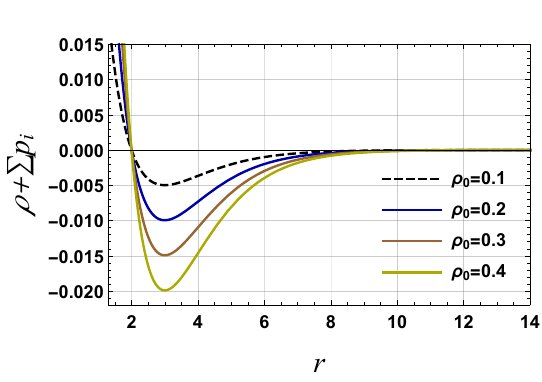} \hspace{2mm}
       \includegraphics[scale=0.89]{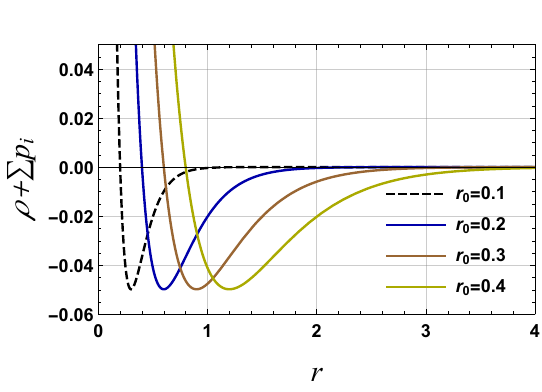} \hspace{2mm}
      	}
      	\caption{The variation of $\rho$ (weak energy condition), $\rho+p_{\theta,\phi} $ (null energy condition), $\rho-\mid p_{\theta,\phi}\mid $ (dominant energy condition), and $\rho+\sum_i p_i$ (strong energy condition) against $r$ for various value of the parameters $\rho_0$ and $r_0$.}
      	\label{ff1}
\end{figure*}

To investigate the curvature structure of the exponential-spheroid-surrounded black hole, we calculate the invariants presented in Eqs. (\ref{r1})–-(\ref{r3}). The Ricci scalar and the Ricci square remain finite at the center, with values $R(0)=-4M_0/r_0^3$ and $R_{\alpha\beta}R^{\alpha\beta}(0)=4M_0^2/r_0^6$, indicating that the halo does not produce a Ricci-type singularity. The Kretschmann scalar exhibits the typical central divergence when \( M_{\rm BH} \neq 0 \), showing leading behavior of \( K \sim 48M_{\rm BH}^2/r^6 \) as \( r \to 0 \). This indicates that while the halo moderates the curvature sourced by matter, it does not remove the tidal singularity associated with the black hole. Conversely, in the limit where \( M_{\rm BH} = 0 \), the Kretschmann scalar becomes finite at the origin. This observation confirms that the singularity is linked to the central black hole rather than the exponential halo itself.

We now examine the weak, null, dominant and strong energy conditions for the anisotropic source fluid generating the exponential spheroid halo.  The energy-momentum tensor may be written in an orthonormal frame as
\begin{equation}\label{Tmunu_halo}
  T^{\mu\nu}
  \;=\;
  \rho\,e_{0}^{\mu}e_{0}^{\nu}
  \;+\;\sum_{i=1}^{3} p_{i}\,e_{i}^{\mu}e_{i}^{\nu},
\end{equation}
where \(\rho\) is the energy density, \(p_{r}\) is the radial (longitudinal) pressure, and \(p_{\theta}=p_{\varphi}\) are the transverse pressures. Thus, regarding the parameter space, the energy density, radial pressure, and transverse pressures are defined as follows:
\begin{align}
  \rho(r)&= - p_{r}(r) = \rho _0 \, e^{-\frac{r}{r_0}} ,\nonumber\\
  \label{rho_halo}\\[6pt] 
p_{\theta}(r) & =  - \frac{\rho _0 \,e^{-\frac{r}{r_0}}}{2r_0}(r-2r_0),\nonumber\\
  \label{pt_halo}
\end{align}
Using these expressions, one checks each pointwise energy condition as follows:

\begin{itemize}
  \item Weak Energy Condition (WEC): \(\rho\ge0\) and \(\rho + P_{i}\ge0\) for \(i=r,\theta,\varphi\).  From \eqref{rho_halo}--\eqref{pt_halo}, 
  where each combination is manifestly nonnegative for $r_0>0,\rho_{0}>0,r>0$.  Hence WEC holds.

  \item Null Energy Condition (NEC): \(\rho + P_{r}\ge0\) and \(\rho + P_{\theta}\ge0\). 
  Both combinations are nonnegative everywhere, so NEC holds.

  \item Dominant Energy Condition (DEC): \(\rho \ge |P_{i}|\) for \(i=r,\theta,\varphi\).  
  Since \(\mathcal{A}(r)<1\), \(\mathcal{A}'(r)<0\), \(\mathcal{A}''(r)>0\) in the halo region, one verifies \(\rho - P_{\theta}\ge0\) for all \(r>0\).  Hence DEC is satisfied.

  \item Strong Energy Condition (SEC): \(\rho + \sum_{i}P_{i} \ge0\), \(\rho + P_{r}\ge0\), \(\rho + P_{\theta}\ge0\).  Together with the NEC combinations shown above, this guarantees SEC holds as well.
\end{itemize}

To gain more insights into the profile of the dark matter halo surrounding the black hole spacetime, Fig. \ref{ff1} clearly shows the variations of related energy conditions' elements, namely, weak, null, dominant, and strong. Concretely, it is observed that the energy density $\rho(r)$ is everywhere nonnegative, while radial pressure fulfils the ansatz $p_r=-\rho$, implying that the radial null energy condition is satisfied. Similarly, it is observed that the combination $\rho-p_{\theta,\phi}$ remains positive with respect to any variation on the parameter space, which means that the dominant energy condition is satisfied. On the other hand, it is shown that the quantity $\rho+\sum_i p_i$ is not positive everywhere. While it is near-zero or slightly positive at small $r$, it drops below zero, indicating a sign change over a finite range of radial coordinates, before asymptotically returning to zero from below at large $r$. Moreover, the depth and extent of this negative region increase as $\rho_0$ rises, and they also expand when $r_0$ is increased. Therefore, it is observed that the SEC is violated within a finite radial interval. Consequently, we conclude that all energy conditions are satisfied, with the exception of the SEC, which is violated within a finite radial interval. 

   \begin{figure*}[htb!]
      	\centering{
      	\includegraphics[scale=0.9]{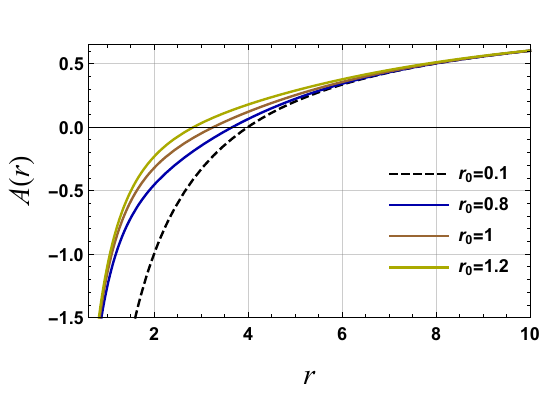}
       \hspace{5mm}
            \includegraphics[scale=0.9]{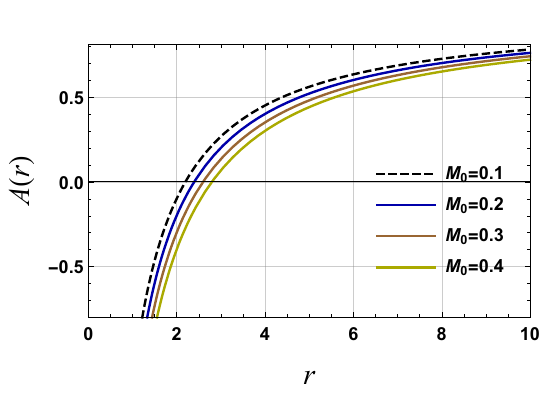}\\
            \includegraphics[scale=0.75]{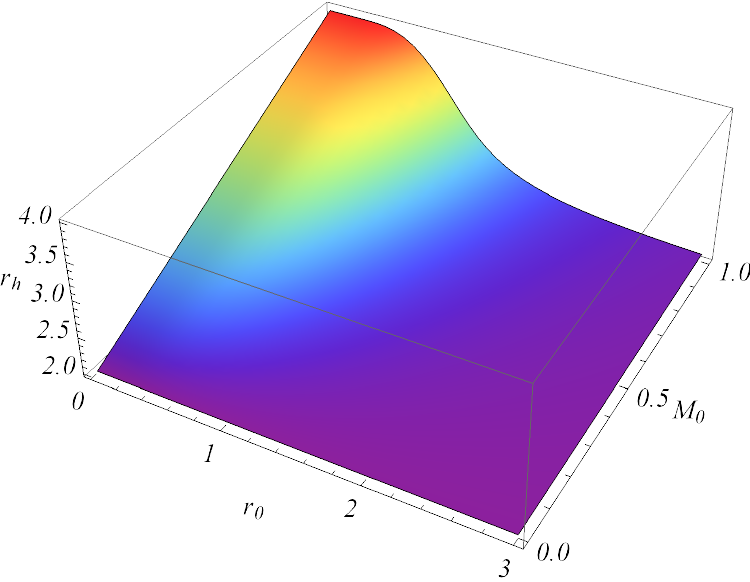}
            }
      	\caption{The metric function $\mathcal{A}(r)$ representation for various value of the parameters $\rho_0$ and $r_0$ with $M=1$ (upper row). The extremal black hole mass with varying $r_0$ and $M_0$ for $M=1$ (lower row).}
      	\label{ff2}
      \end{figure*}

\section{Black hole shadow silhouettes with EHT constraints (Revised by Soroush) }\label{sec2}

\begin{figure}[!htb]
\center{
	\includegraphics[width=7cm]{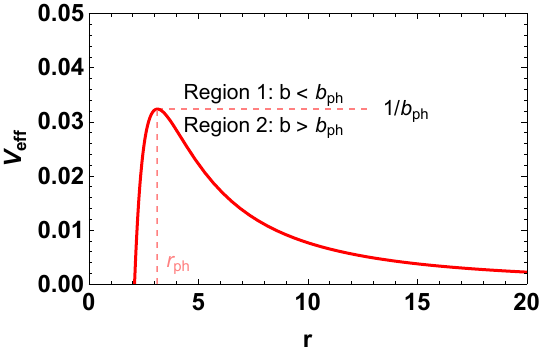}}
\caption{Plots of the effective potential $V_{\text{eff}}$ and and impact parameter $b_{\rm ph}$ as functions of the radial coordinate $r$.}
\label{veff}
\end{figure}

\begin{figure*}
		\includegraphics[width=8cm]{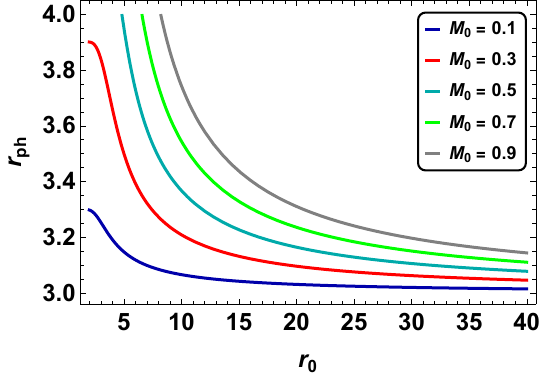}
		\includegraphics[width=8cm]{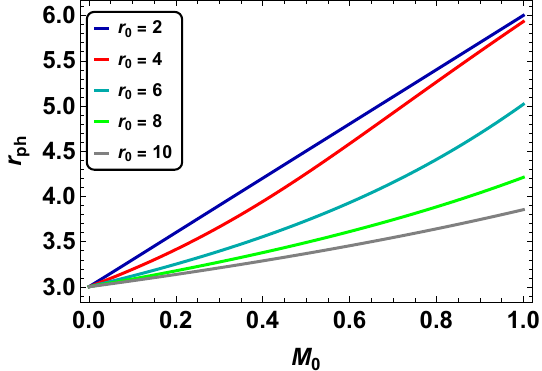}
		\includegraphics[width=8cm]{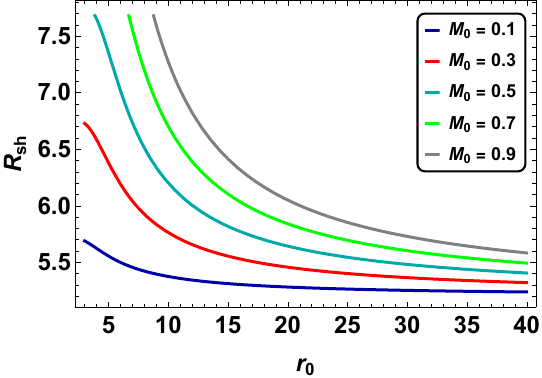}
		\includegraphics[width=8cm]{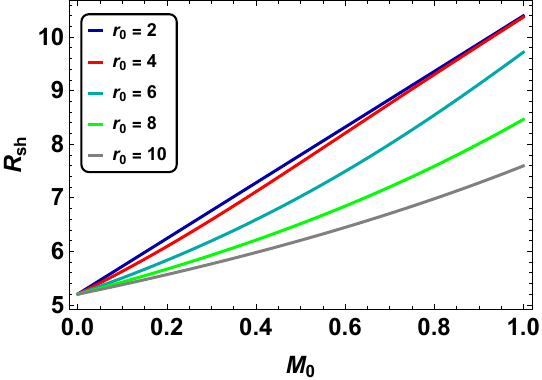}
	\caption{The upper panels display the photon-sphere radius $r_{\rm ph}$ as a function of the deviation parameters $M_{0}$ and $r_{0}$, while the lower panels show the corresponding shadow radius $R_{\rm sh}$ as a function of the same parameters.}
	\label{rphrsh}
\end{figure*}

In this section, we derive the equations governing the null geodesic motion around a black hole embedded in an exponential spheroidal dark matter halo. The trajectories of photons are obtained from the corresponding Euler-Lagrange equations, which take the following form:
\begin{eqnarray}\label{EulerLagrangeEq}
\frac{d}{d\tau}\Big(\frac{\partial \mathcal{L}}{\partial \dot{x}^\mu}\Big)-\frac{\partial \mathcal{L}}{\partial x^\mu}=0,
\end{eqnarray}
 in which $\dot{x}^\mu = dx^\mu/d\tau$ denotes the four-velocity of the photon, and $\tau$ is the affine parameter. For a static, spherically symmetric Schwarzschild-type black hole spacetime, the Lagrangian describing the photon’s motion can be written as follows:
\begin{eqnarray}
	\nonumber
	\mathcal{L}=\frac{1}{2}g_{\mu\nu}\dot{x}^\mu\dot{x}^\nu
	=\frac{1}{2}\big[-f(r)\dot{t}^2+\frac{1}{f(r)}\dot{r}^2+r^2(\dot{\theta}^2+\sin^2\theta\dot{\phi}^2)\big].
\end{eqnarray}
 Because $\partial_t$ and $\partial_\phi$ are Killing vector fields of the Schwarzschild-type black hole spacetime embedded in a dark matter halo, the photon’s motion admits two conserved quantities: the energy $E$ and the azimuthal component of the angular momentum $L_z$, given by
\begin{eqnarray}
	E\equiv-\frac{\partial \mathcal{L}}{\partial \dot{t}}=f(r)\dot{t},\quad L_z=\frac{\partial \mathcal{L}}{\partial \mathcal{\dot{\phi}}}=r^2 \sin^2\theta \dot{\phi}.
\end{eqnarray}
Furthermore, the spherical symmetry of the spacetime allows us to restrict the photon motion to the equatorial plane $\theta = \pi/2$ without loss of generality. By setting $\mathcal{L} = 0$ for null geodesics and introducing the impact parameter $b = L_{z}/E$, one then obtains three first-order differential equations governing the photon motion:
\begin{eqnarray}
	\dot{t}=\frac{E}{f(r)}, \quad \dot{\phi}=\pm\frac{L_{z}}{r^2}
	\label{EqPhi},~~~
	\dot{r}^2=\frac{1}{L_{z}^{2}}\left(\frac{1}{b^{2}}-V_{\text{eff}}(r)\right),
	\label{Eq-r-motion}
\end{eqnarray}
where, the $\pm$ sign in \eqref{EqPhi} corresponds to the clockwise and counterclockwise directions of the photon trajectories, respectively. The fate of a photon, determined by its impact parameter $b_{\rm ph}$, is subsequently governed by the radial geodesic equation \cite{CunhaGRG2018}. Here, the corresponding effective potential can be expressed as
\begin{eqnarray}
	V_{\text{eff}}(r)=\frac{f(r)}{r^2}.
	\label{eq:Veff}
\end{eqnarray}
The asymptotic flatness of the metric at spatial infinity requires that $V_{\text{eff}}(r)$ decreases as $1/r^{2}$ for $r \to \infty$. At the event horizon $r_{+}$, we have $V_{\text{eff}}(r_{+}) = 0$ since $f(r_{+}) = 0$. Consequently, $V_{\text{eff}}(r)$ must possess at least one maximum between $r_{+}$ and spatial infinity.
Specifically, when the impact parameter satisfies $\frac{1}{b^{2}} = V_{\text{eff}}(r_{0})$ at a certain radius $r = r_{0}$, the condition $\dot{r} = 0$ holds, indicating that the photon is deflected at its minimum radial distance $r_{0}$ from the black hole. The critical impact parameter $b_{\text{ph}}$ is then obtained from the condition
$\dot{r} = 0$ at the maximum of $V_{\text{eff}}$ \cite{Chandrasekhar,StuchlikEPJC2019,StuchlikApJ2019}
\begin{eqnarray}
	b_{\rm ph}&=&\frac{1}{\sqrt{V_{\text{eff}}(r_{\rm ph})}},\\  0&=&V'_{\text{eff}}(r_{\rm ph})=\frac{r_{\rm ph}f'(r_{\rm ph})-2f(r_{\rm ph})}{r_{\rm ph}^{3}}, .
	\label{eq:rph}
\end{eqnarray}
where $r_{\rm ph}$ is the photon-sphere radius satisfying $V_{\text{eff}}''(r_{\rm ph}) < 0$.
Thus, geodesic trajectories with the critical impact parameter $b_{\rm{ph}}$ correspond to the photon sphere of radius $r_{\rm{ph}}$ \cite{CapozzielloJCAP2023,CapozzielloPoDU2025,NietoPoDU2025}. 
As shown in Fig. \ref{veff}, the orbits of these photons are inherently unstable, such that a small radial perturbation causes the photon either to escape to infinity $b > b_{\rm{ph}}$ or to be captured by the black hole $b < b_{\rm{ph}}$ \cite{Meng-KuangEPJC2024,Meng-KuangPRD2023}. 
 For a distant observer viewing a black hole described by an asymptotically flat metric, the shadow radius reduces to $R_{\rm sh} = b_{\rm ph}$.
  Using the radius of the photon sphere $r_{\rm ph}$, one can then readily determine the corresponding critical impact parameter $b_{\rm ph}$, which defines the shadow radius $R_{\rm sh}$ of the black hole as \cite{KonoplyaPLB2019}
   \begin{equation}\label{shadowRadius}
  	R_{\rm sh} = b_{\rm ph} = \frac{r_{\rm ph}}{\sqrt{f(r_{\rm ph})}}.
  \end{equation}
However, this relation does not always hold. For instance, in non-asymptotically flat spacetimes -- such as the Kottler (Schwarzschild-de Sitter) spacetime -- one finds that $R_{\rm sh} \neq b_{\rm ph}$.
The influence of the deviation parameters on $r_{\rm ph}$ and $b_{\rm ph}$ is illustrated in the first and second rows of Fig. \ref{rphrsh}, respectively. The plots show that, for a fixed value of $r_{0}$, increasing $M_{0}$ leads to a larger photon-sphere radius $r_{\rm ph}$ and a correspondingly larger shadow radius. In contrast, for a fixed value of $M_{0}$, increasing $r_{0}$ decreases both the photon-sphere radius $r_{\rm ph}$ and the shadow radius.
The Event Horizon Telescope (EHT) observations of M87* and Sgr A* impose valid constraints on the parameters of the exponential spheroidal dark matter halo. According to the observational data reported in Refs. \cite{EHTL1,EHTL5,EHTL6,EHTL12,EHTL17,Do,GRAVITY,GRAVITY2020,Kocherlakota,Vagnozzi}, the shadow sizes of M87* and Sgr A* are constrained within the intervals $[4.26,6.03]$ and $[4.55,5.22]$ at the $1\sigma$ confidence level (CL), respectively. At the $2\sigma$ confidence level, the corresponding ranges are $[3.38,6.91]$ for M87* and $[4.21,5.56]$ for Sgr A* \cite{ZareJCAP2024,Zare2025,ZarePLB2024,SekhmaniJHEA2025,Meng2025}. 

\begin{figure}[!htb]
\center{
\includegraphics[width=6.1 cm]{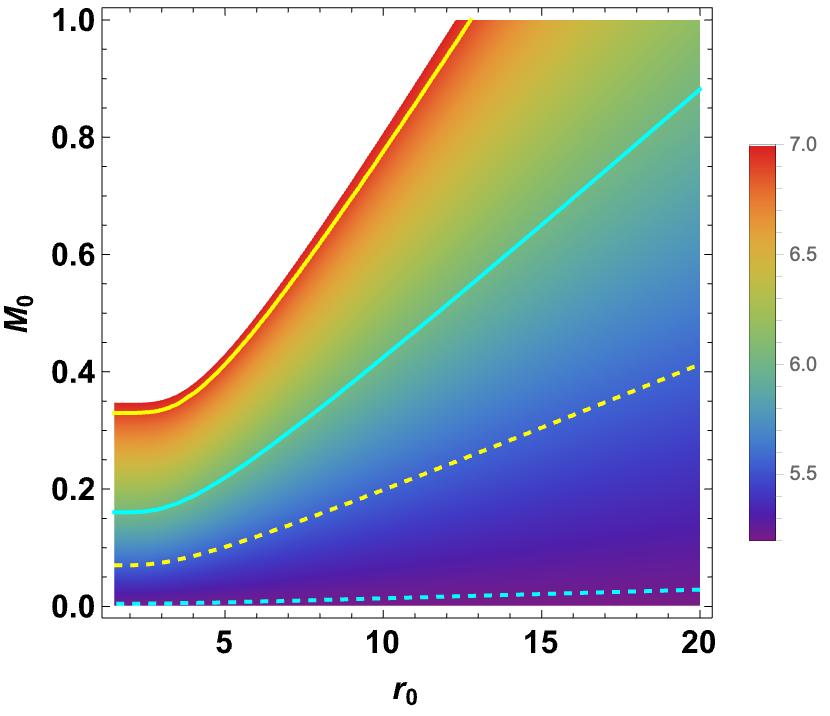}}
\caption{ Density of the shadow radius as a function of $M_{0}$ and $r_{0}$. The colored contours indicate the upper limits of the allowed regions at the $1\sigma$ (cyan) and $2\sigma$ (yellow) CLs, derived from EHT observations of Sgr A* (dashed curves) and M87* (solid curves).}
\label{solutionscases}
\end{figure}

Figure \ref{solutionscases} displays the dependence of the black hole shadow radius $R_{\rm sh}$ on the exponential dark matter halo parameters $M_0$ and $r_0$, together with the $1\sigma$ (cyan) and $2\sigma$ (yellow) confidence regions inferred from the EHT measurements of M87* (solid) and Sgr A* (dashed). 
Since $R_{\rm sh}$ decreases monotonically with $r_{0}$ and increases monotonically with $M_{0}$ (i.e., $\partial R_{\rm sh}/\partial r_{0}<0$ and $\partial R_{\rm sh}/\partial M_{0}>0$), the resulting contours define the upper boundaries of the parameter space compatible with the observed shadow sizes. Consequently, all parameter combinations located below and to the right of a given contour are consistent with the EHT measurements, whereas points above or to the left would predict shadow radii exceeding the observational bounds. 
Consequently, the constraints are one-sided (upper) when projected onto each parameter axis. 
For M87*, the figure indicates that the 1$\sigma$ contour confines the parameters roughly to $r_0\lesssim 20$ and $M_0\lesssim 0.9$, while the $2\sigma$ contour extends to $r_0\lesssim 13$ and $M_0\lesssim 1$. For Sgr A*, the corresponding limits are slightly smaller, with $r_0\lesssim 20$, $M_0 \lesssim 0.04 \, (1 \sigma)$ and $r_0\lesssim 20, M_0\lesssim 0.4\, (2 \sigma)$.
These values should be read as approximate upper bounds extracted from the contour map; smaller densities and core radii remain fully compatible with the data, since the pure-Schwarzschild (no-halo) case already yields a shadow within the EHT range. Physically, this means that any exponential dark matter halo surrounding M87* or Sgr A* must be diffuse and compact, contributing only a minor correction to the central gravitational potential near the photon sphere. 

\section{Quasinormal Modes}
\label{sec:qnm}

In this section we study the QNMs of the black hole solution. QNMs characterize the linear response of a black hole to perturbations by test fields and appear as complex frequencies whose real part sets the oscillation frequency and whose imaginary part gives the exponential decay rate. We analyze the frequency-domain QNMs for scalar perturbation using the 6$^{\rm th}$ order Pad\'e-averaged WKB method. We begin with a succinct review of the Pad\'e-averaged WKB approach.

\subsection{Pad\'e-averaged WKB method for QNMs}
\label{sec:padeqnm}
The semi-analytic WKB approach for quasinormal modes (QNMs) originated with Mashhoon \cite{mashhoon}, who correlated the effective potential with an inverse P{\"o}schl--Teller form. Schutz and Will \cite{schutz1985black} later refined this method by conducting a WKB expansion around the peak of the effective potential and aligning solutions in the asymptotic regions. Iyer and Will \cite{iyer1987black} expanded the WKB method to third order, which enhanced the accuracy for the fundamental mode when \(\ell\gg n\). Subsequent advancements have extended the method to even higher orders, reaching up to sixth order and beyond \cite{Konoplya6thOrder,konoplya2019higher, 2021EPJC...81..475M, 2025JHEAp..4700389S, 2025PDU....4801865A, 2025JHEAp..45..200S, 2026PDU....5202358S}. However, the standard WKB series can be asymptotic, and its termwise convergence is not assured, particularly for overtones where \(n\gtrsim\ell\).

To improve convergence one constructs Pad\'e approximants of the WKB series \cite{matyjasekopalaWKB,konoplya2019higher}. For a Schr\"odinger-like wave equation
\begin{equation}\label{eq:wave_standard}
\frac{d^2\Psi}{dx^2} + \big(\omega^2 - V(x)\big)\Psi = 0,
\end{equation}
the WKB expansion gives a formal expression for \(\omega^2\) of the form
\begin{equation}\label{eq:omega_wkb}
\omega^2 = V_0 + A_2(\mathcal{K}^2) + A_4(\mathcal{K}^2) + A_6(\mathcal{K}^2)+\ldots
    - i\,\mathcal{K}\sqrt{-2V_2}\Big(1 + A_3(\mathcal{K}^2)+A_5(\mathcal{K}^2)+\ldots\Big),
\end{equation}
where \(V_0\) and \(V_2\) are the value and the second derivative of the potential at its maximum, the quantities \(A_k\) are known analytic corrections that depend on derivatives of the potential, and
\begin{equation}
\mathcal{K} \;=\; n+\tfrac{1}{2}, \qquad n=0,1,2,\dots
\end{equation}
is the usual half–integer overtone parameter for \(Re(\omega)>0\) (the sign choice for \(Re(\omega)<0\) follows by symmetry).

One introduces an order parameter \(\epsilon\) and reorganises the right-hand side of \eqref{eq:omega_wkb} as a polynomial \(P_k(\epsilon)\) in powers of \(\epsilon\). The Pad\'e approximant \(P_{\tilde n/\tilde m}(\epsilon)\) is then constructed as the rational function
\begin{equation}
P_{\tilde n/\tilde m}(\epsilon)=\frac{Q_0+Q_1\epsilon+\cdots+Q_{\tilde n}\epsilon^{\tilde n}}{R_0+R_1\epsilon+\cdots+R_{\tilde m}\epsilon^{\tilde m}},\qquad
\tilde n+\tilde m=k,
\end{equation}
which typically yields a much better behaved approximation than the truncated polynomial. In practice one computes Pad\'e approximants at several orders (up to the maximal available order) and selects the result that minimises an internal error estimate. A convenient error proxy at order \(k\) is
\begin{equation}
\Delta_k \;=\; \frac{|\omega_{k+1}-\omega_{k-1}|}{2},
\end{equation}
where \(\omega_j\) denotes the frequency obtained at WKB (Pad\'e) order \(j\). The best estimate is taken at the order (and Pad\'e index) for which \(\Delta_k\) is smallest.

  \begin{figure*}[!htp]
      	\centering{
       \includegraphics[scale=0.93]{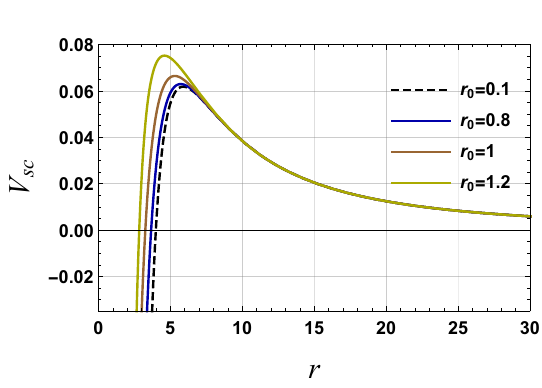} \hspace{2mm}
      	\includegraphics[scale=0.93]{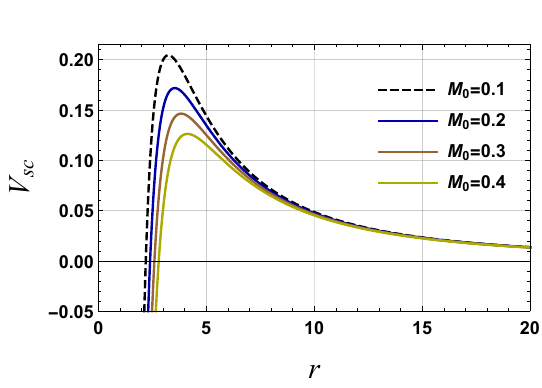} \hspace{2mm}
      	}
      	\caption{The scalar potential variation for various value of $r_0$ and $M_0$ with $M=1$ and $\ell=2$.}
      	\label{pot}
      \end{figure*}

\begin{table}[ht!]
\caption{The table presents the QNMs' frequencies for scalar perturbations, calculated using a 6th-order Padé method. It displays results for various values of the black hole parameters $r_0$ and $M_0$, alongside multiple values of the overtone $n$ and the multipole quantum number $\ell$, while maintaining a fixed value of $M=1$.}
\centering
\label{t1}
\begin{tabular}{|c|c|c|c|c|}
\hline
\multicolumn{5}{|c|}{$M_0=0.5$}\\
\hline
$r_0=0.1$ & $n=0$ & $n=1$ & $n=2$ & $n=3$\\
\hline
$\ell=1$ & $0.195287 - 0.0651072i$ & $ 0.176303 - 0.20434i$ & $ 0.152568 - 0.361453i$ & $0.136251 - 0.532807i$\\
\hline
$\ell=2$ & $0.322429 - 0.0645061  i$ & $0.309231 - 0.197084  i$ & $0.286886 - 0.339111  i$ & $0.261663 - 0.493085   i$\\
\hline
$\ell=3$ & $0.450244 - 0.0643331i$ & $0.440447 - 0.194858 i$ & $0.42239 - 0.330669 i$ & $0.398888 - 0.474205 i$\\
\hline
\hline
$r_0=1$ & $n=0$ & $n=1$ & $n=2$ & $n=3$\\
\hline
$\ell=1$ & $0.213868 - 0.0592098 i$ & $0.212177 - 0.184029 i$ & $0.265087 - 0.355368 i$ & $0.369557 - 0.339162 i$\\
\hline
$\ell=2$ & $0.351706 - 0.0583907  i$ & $0.34736 - 0.177048 i$ & $0.347481 - 0.300142 i$ & $0.369376 - 0.410779 i$\\
\hline
$\ell=3$ & $0.490465 - 0.0582006  i$ & $0.486527 - 0.17545 i$ & $0.481353 - 0.294578 i$ & $0.480335 - 0.413101 i$\\
\hline
\hline
\hline
\multicolumn{5}{|c|}{$M_0=1$}\\
\hline
$r_0=0.1$ & $n=0$ & $n=1$ & $n=2$ & $n=3$\\
\hline
$\ell=1$ & $0.146484 - 0.0488186  i$ & $0.131756 - 0.152741  i$ & $0.107701 - 0.269532 i$ & $0.0782681 - 0.398567 i$\\
\hline
$\ell=2$ & $0.242416 - 0.0484758 i$ & $0.235466 - 0.147i$ & $0.224318 - 0.247172 i$ & $0.210065 - 0.346854  i$\\
\hline
$\ell=3$ & $0.337953 - 0.048309  i$ & $0.332088 - 0.146005  i$ & $0.32234 - 0.245483 i$ & $0.310442 - 0.345542 i$\\
\hline
\hline
$r_0=1$ & $n=0$ & $n=1$ & $n=2$ & $n=3$\\
\hline
$\ell=1$ & $0.154414 - 0.0411252 i$ & $0.147992 - 0.128098  i$ & $0.142305 - 0.230074 i$ & $0.148503 - 0.340001  i$\\
\hline
$\ell=2$ & $0.253621 - 0.040586 i$ & $0.250197 - 0.123601i$ & $0.244805 - 0.212586 i$ & $0.241141 - 0.308396 i$\\
\hline
$\ell=3$ & $0.353491 - 0.0403453  i$ & $ 0.351116 - 0.122065i$ & $0.346998 - 0.206903 i$ & $0.342544 - 0.296255  i$\\
\hline
\end{tabular}
\end{table}

\begin{table}[ht!]
\caption{The table presents the QNMs' frequencies for scalar perturbations, calculated using a 6th-order WKB method. It displays results for various values of the black hole parameters $r_0$ and $M_0$, alongside multiple values of the overtone $n$ and the multipole quantum number $\ell$, while maintaining a fixed value of $M=1$.}
\centering
\label{t2}
\begin{tabular}{|c|c|c|c|c|}
\hline
\multicolumn{5}{|c|}{$M_0=0.5$}\\
\hline
$r_0=0.1$ & $n=0$ & $n=1$ & $n=2$ & $n=3$\\
\hline
$\ell=1$ & $0.195273 - 0.0651744 i$ & $ 0.176314 - 0.204345 i$ & $ 0.15401 - 0.361443 i$ & $0.262138 - 0.493256   i$\\
\hline
$\ell=2$ & $0.322428 - 0.0645107 i$ & $0.309231 - 0.197085i$ & $0.286924 - 0.339133i$ & $0.262138 - 0.493256 i$\\
\hline
$\ell=3$ & $0.450244 - 0.0643337i$ & $0.440447 - 0.194858 i$ & $0.422394 - 0.330674i$ & $0.398954 - 0.474255 i$\\
\hline
\hline
$r_0=1$ & $n=0$ & $n=1$ & $n=2$ & $n=3$\\
\hline
$\ell=1$ & $0.21209 - 0.0614151 i$ & $0.209741 - 0.202939i$ & $0.25196 - 0.355756    i$ & $0.406468 - 0.466195  i$\\
\hline
$\ell=2$ & $0.351647 - 0.058495 i$ & $0.347262 - 0.177837 i$ & $ 0.347002 - 0.300061i$ & $0.366631 - 0.41435 i$\\
\hline
$\ell=3$ & $ 0.490459 - 0.0582143  i$ & $ 0.486522 - 0.175517 i$ & $ 0.481301 - 0.294544  i$ & $ 0.480116 - 0.413241i $\\
\hline
\hline
\hline
\multicolumn{5}{|c|}{$M_0=1$}\\
\hline
$r_0=0.1$ & $n=0$ & $n=1$ & $n=2$ & $n=3$\\
\hline
$\ell=1$ & $ 0.226113 - 0.0316602i$ & $ 0.813779 - 0.0249034 i$ & $ 2.34138 - 0.0133724i$ & $ 5.28391 - 0.0083541 i$\\
\hline
$\ell=2$ & $ 0.269535 - 0.0476597 i$ & $ 0.559234 - 0.0838408i$ & $ 1.48317 - 0.0840334i$ & $ 3.31769 - 0.10171i$\\
\hline
$\ell=3$ & $ 0.343808 - 0.0488052i$ & $ 0.427741 - 0.12537i$ & $ 0.836178 - 0.129435i$ & $ 1.79236 - 0.122053i$\\
\hline
\hline
$r_0=1$ & $n=0$ & $n=1$ & $n=2$ & $n=3$\\
\hline
$\ell=1$ & $ 0.154856 - 0.04001 i$ & $ 0.147836 - 0.123301i$ & $ 0.141109 - 0.216534  i$ & $ 0.153393 - 0.312545i$\\
\hline
$\ell=2$ & $ 0.253657 - 0.0404917 i$ & $0.250154 - 0.123154 i$ & $0.244788 - 0.210502  i$ & $0.241389 - 0.303368 i$\\
\hline
$\ell=3$ & $0.353496 - 0.0403313i$ & $ 0.351112 - 0.121999i$ & $ 0.347003 - 0.206554 i$ & $ 0.342602 - 0.295318i$\\
\hline
\end{tabular}
\end{table}

\begin{table}[ht!]
\centering
\caption{Comparison of the numerical stability and physical consistency between the standard 6th-order WKB method and the 6th-order Padé resummation technique for calculating scalar quasinormal modes in the ESM-surrounding field black hole spacetime.}
\label{tab:method_comparison}
\renewcommand{\arraystretch}{1.3} 
\begin{tabular}{|p{3.5cm}|p{6cm}|p{5cm}|}
\hline
Regime & Observation & Interpretation \\
\hline
$n=0$, moderate $\ell$ & Padé and WKB nearly coincide, e.g., $M_0=0.5$, $r_0=1$, $\ell=3$: $0.490465 - 0.0582006i$ vs. $0.490459 - 0.0582143i$. & Both methods are reliable for the fundamental mode. \\
\hline
Low overtones, $n < \ell$ & Agreement remains good in most cases. & Standard WKB still works in its expected domain. \\
\hline
Higher overtones, $n \ge \ell$ & WKB begins to drift away from Padé. & The asymptotic WKB expansion is losing convergence. \\
\hline
Strong halo, compact core \newline ($M_0=1$, $r_0=0.1$) & WKB becomes non-smooth and gives very large real parts, e.g., $\ell=1$, $n=3$: $5.28391 - 0.0083541i$, while Padé stays smooth at $0.0782681 - 0.398567i$. & This is the clearest sign that Padé is the more trustworthy method in the problematic region. \\
\hline
\end{tabular}
\end{table}

\begin{figure}[!htb]
     \centering
{\includegraphics[scale=0.63]{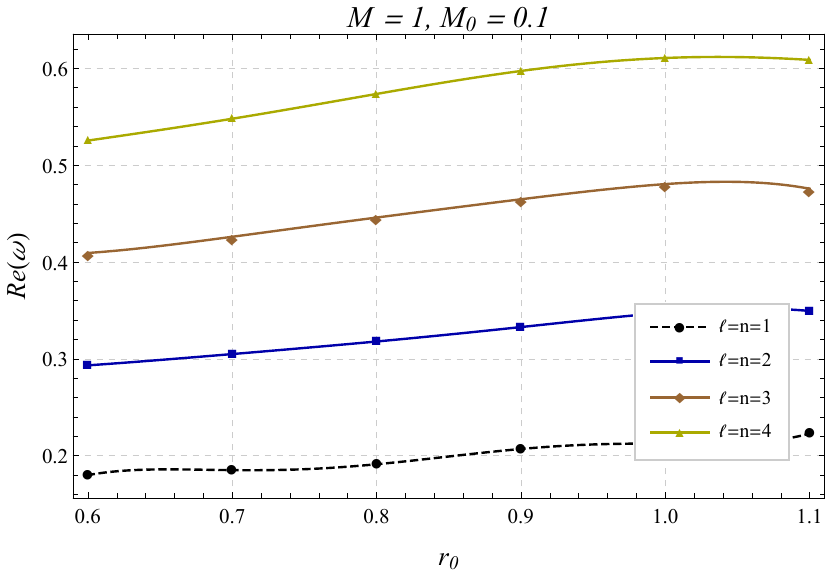}}
         \label{fig:potvarl}
{\includegraphics[scale=0.633]{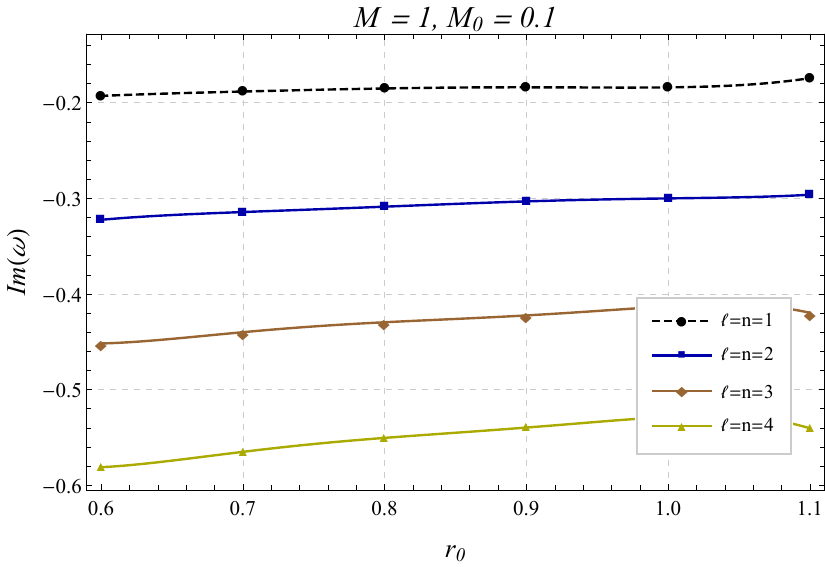}}
         	\caption{The scalar QNMs real frequency $Re(\omega)$ and the damping rate $Im(\omega)$ behavior with respect to $r_0$ for various overtones $n$ and multipole number $\ell$.}\label{qqn}
\end{figure}

\subsection{Massless scalar perturbations}
We next specialise to massless scalar perturbations obeying the Klein--Gordon equation
\begin{equation}\label{scalar_KG_repeat}
\square\Phi \;=\; \frac{1}{\sqrt{-g}}\partial_\mu\!\big(\sqrt{-g}\,g^{\mu\nu}\partial_\nu\Phi\big)=0.
\end{equation}
Working to linear order in the perturbation and neglecting backreaction on the background geometry, the metric may be taken in the static, spherically symmetric form
\begin{equation}
ds^2 = -|g_{tt}(r)|\,dt^2 + g_{rr}(r)\,dr^2 + r^2 d\Omega^2,
\end{equation}
so that the scalar field admits the standard decomposition
\begin{equation}
\Phi(t,r,\theta,\varphi) = \frac{1}{r}\sum_{\ell,m}\psi_{\ell}(t,r)\,Y_{\ell m}(\theta,\varphi).
\end{equation}
Introducing the tortoise coordinate \(r_*\) defined by
\begin{equation}\label{tortoise_correct}
\frac{dr_*}{dr} = \sqrt{\frac{g_{rr}}{|g_{tt}|}},
\end{equation}
and assuming the harmonic time dependence \(\psi_\ell(t,r)=\psi_\ell(r)e^{-i\omega t}\), the radial equation reduces to the standard Schr\"odinger–like form
\begin{equation}\label{radial_schrod}
\frac{d^2\psi_\ell}{dr_*^2} + \big(\omega^2 - V_{sc}(r)\big)\psi_\ell = 0,
\end{equation}
where the effective potential for scalar perturbations reads
\begin{equation}\label{Vs_correct}
V_{sc}(r) \;=\; |g_{tt}(r)|\left[\frac{\ell(\ell+1)}{r^2}
    +\frac{1}{r\sqrt{|g_{tt}|/g_{rr}}}\frac{d}{dr}\!\Big(\sqrt{\frac{|g_{tt}|}{g_{rr}}}\,\Big)\right].
\end{equation}
The potential \eqref{Vs_correct} is real, positive outside the horizon for \(\ell\geq0\) and typically has a single maximum; this peak is the input for the WKB-Pad\'{e} procedure described above.

To deepen our understanding of the scalar perturbation process, Fig. \ref{pot} illustrates the behavior specific to the scalar potential; in other words, Figure 6 shows the origin, at the potential level, of the decay behavior and the greybody. So, this is the key barrier liable to control quasi-normal scalar modes. Thus, for both panels, the potential exhibits a single peak beyond the horizon after having decayed to zero at large radii. More specifically, for a fixed $M_0$, an increasing variation in the halo parameter $r_0$ leads to a slight raising and a slight broadening of the barrier, shifting the peak outward. On the other hand, an increase of $M_0$ with a fixed value of $r_0$ involves a decrease of the barrier peak with a changing of its location more significantly. This implies that the halo attenuates the scattering barrier in a way that affects both the quasi-normal spectrum and the transmission through the geometry.

Tables \ref{t1} and \ref{t2} show, respectively, the scalar frequencies of the quasi-normal modes obtained using the 6th-order WKB method with Pad\'{e} approximants and the conventional 6th-order WKB method. To evaluate the numerical robustness of the spectrum, it is useful to compare the two sets of results under different parameter regimes.

For the fundamental modes ($n=0$) and the low harmonics such as $n < \ell$, the two methods converge quite closely. For example, in a spread halo background ($M_0=0.5$, $r_0=1$) for the mode $\ell=3$, $n=0$, the Pad\'{e} method results in $0.490465 - 0.0582006i$, which agrees almost perfectly with the standard WKB result of $0.490459 – 0.0582143i$. This outstanding agreement demonstrates that the underlying potential is sufficiently consistent and that both methods are reliable in this regime.

In order to determine the numerical robustness of the spectrum of scalar quasi-normal modes, we compared the results of the WKB method according to the 6th-order Pad\'{e} approximation with those of the standard 6th-order WKB frequencies (see Tab. \ref{tab:method_comparison}). The two methods converge very well for the fundamental modes and for the low overtones such that $n < \ell$, which implies that the background potential is sufficiently smooth in this regime. However, as the overtone number increases and approaches, or even exceeds, the multipole number, the simple WKB approximation gradually ceases to converge. This discrepancy is particularly noticeable in the system characterized by a strong halo and a compact core ($M_0=1, r_0=0.1$), where the WKB frequencies exhibit an irregular and physically unrealistic growth in $\text{Re}(\omega)$, whilst the frequencies obtained using the Pad\'{e} approximation remain regular and are consistently damped. We therefore consider that the spectrum given by the Pad\'{e} approximation is the most reliable quantitative result and use the standard WKB spectrum only as a reference for low-energy modes.

Fig. \ref{qqn} shows for various sets of equal overtone and multipole numbers, the variation of the real part $\mathrm{Re}(\omega)$ and the damping rate $\mathrm{Im}(\omega)$ against the parameter $r_0$ for the fixed sets of $M_0=1$ and $M=1$. Thus, the real part $\mathrm{Re}(\omega)$ and the damping rate $\mathrm{Im}(\omega)$ both depend on the halo scale $r_0$, the overtone number $n$, and the multipole number $\ell$. The depicted curves illustrate the expected hierarchy: higher values of $\ell$ correspond to higher oscillation frequencies, while the higher overtones are attenuated to a greater extent. The dependence on $r_0$ is relatively weak in the displayed plot, but the tendency is still noticeable: the damping decreases as $r_0$ increases, resulting in oscillations of longer-lived oscillations. 

\section{Greybody bounds}\label{GF}
In this section, we examine the bounds of greybody factors. Our analysis in the previous section, which utilized the WKB method for quasinormal modes (QNMs), underscores the considerable impact of black hole model parameters on the QNM spectrum. We will now explore scalar perturbations in relation to greybody factors and assess how these model parameters influence the established bounds through an analytical approach.  Analytical techniques for predicting rigorous limits of greybody factors were first developed by Visser \cite{Visser:1998ke} and subsequently refined by Boonserm and Visser \cite{Boonserm:2008zg}. Further studies by Boonserm et al. \cite{Boonserm:2017qcq}, Yang et al. \cite{Yang:2022ifo}, Gray et al. \cite{Gray:2015xig}, Ngampitipan et al. \cite{Ngampitipan:2012dq}, and others \cite{Chowdhury:2020bdi,Miao:2017jtr,Liu:2021xfs,Barman:2019vst,Xu:2019krv,Boonserm:2017qcq} have gone deeper into these limits. Our work expands this investigation by examining a black hole solution with dark matter halos, thereby deepening our understanding of gray body factors across different contexts.

We examine the constraints on grey-body factors associated with black holes that are surrounded by a dark matter halos, focusing specifically on massless scalar perturbations. To facilitate this analysis, we delve into the Klein-Gordon equation that governs the behaviour of the massless scalar field, as discussed in the preceding section. We turn to consider the reduced effective potential, $V_{sc}(r)$, which is given by:
\begin{equation}
    V_{sc}(r) = \frac{\ell(\ell + 1)\mathcal{A}(r)}{r^{2}} + \frac{\mathcal{A}(r)\mathcal{A}'(r)}{r}.\label{poten}
\end{equation}

Next, we utilize the effective potential identified earlier to investigate the lower bound on the greybody factor relevant to our black hole solution. In light of the work by Visser \cite{Visser:1998ke} and Boonserm and Visser \cite{Boonserm:2008zg}, the appropriate method for establishing this stringent limit is as follows:
\begin{equation} \label{bound}
A_g^2 \geq \operatorname{sech}^{2}\left(\frac{1}{2 \omega} \int_{-\infty}^{\infty}\left|V_{sc}\right| \frac{d r}{\mathcal{A}(r)} \right),
\end{equation}
where $A_g^2$ denotes the transmission coefficient $T$ in this context.
    \begin{figure*}[t!]
      	\centering{\includegraphics[scale=0.9]{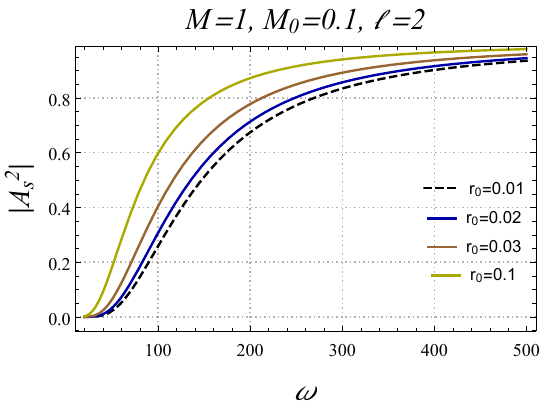}\hspace{5mm}
      	\includegraphics[scale=0.9]{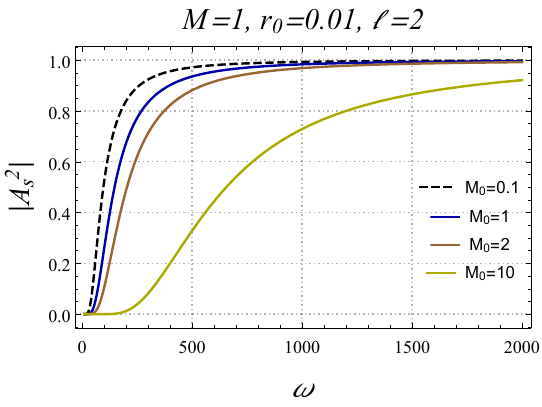} 
       
       }
      	\caption{Rigorous lower bounds  on greybody factors of scalar massless field for various values of the parameters pair $(r_0,M_0)$.}
      	\label{gf}
      \end{figure*}

Additionally, to account for the influence of the cosmological horizon, we adjust the boundary conditions as demonstrated by Boonserm et al. \cite{Boonserm:2019mon}. The modified boundary conditions are:
\begin{equation}
A_g^2 \geq A^2_{s}=\operatorname{sech}^{2}\left(\frac{1}{2 \omega} \int_{r_{H}}^{R_{H}} \frac{|V_{sc}|}{\mathcal{A}(r)} d r\right)=\operatorname{sech}^{2}\left(\frac{A_{l}}{2 \omega}\right),
\end{equation}
where we define
 \begin{equation}
A_{l}=\int_{r_{H}}^{R_{H}} \frac{|V_{sc}|}{\mathcal{A}(r)} d r=\int_{r_{H}}^{R_{H}}\left|\frac{\ell(\ell+1)}{r^{2}}+\frac{\mathcal{A}^{\prime}(r)}{r}\right| d r.
\end{equation}

Here, $r_H$ is the event horizon and $R_H$ is the cosmological horizon of the BH. This specification supplies a rigorous lower bound on the grey-body factors relative to the BH solution.

Fig. \ref{gf} illustrates the rigorous lower bounds on the greybody factors. Thus, the transmission bound increases monotonically with frequency and approaches unity for large values of $\omega$, which corresponds to the expected behavior of a physical greybody factor bound. The halo parameters reduce the bound in the low and mid-frequency ranges: a higher value of $M_0$ results in significantly greater damping, and increasing $r_0$ also shifts the transmission curve downwards for a given $\omega$. The halo therefore makes it more difficult for scalar modes to pass through the effective barrier. This figure thus complements Fig. \ref{pot}: a stronger barrier results in a lower lower bound for transmission at low frequencies.

\section{Conclusion}
\noindent In this work, we constructed and analyzed a Schwarzschild-like black hole immersed in an exponential dark matter halo, using a simple but analytically tractable density profile characterized by the amplitude \(\rho_0\) and the scale radius \(r_0\). The resulting spacetime is asymptotically flat, reduces smoothly to the Schwarzschild geometry in the absence of the halo, and provides a clean framework for studying how environmental matter modifies strong-gravity physics. Our analysis shows that the halo affects the near-horizon structure in a controlled manner and can shift the extremal horizon configuration, with the halo contribution raising the threshold for black-hole formation. 

\noindent The curvature analysis reveals that the halo regularizes the Ricci sector near the origin, since the Ricci scalar and Ricci square remain finite at \(r\to 0\), while the Kretschmann scalar still diverges when the central black-hole mass is present. This confirms that the tidal singularity remains associated with the black-hole core rather than with the halo itself. The energy-condition study further indicates that the weak, null, and dominant energy conditions are satisfied, whereas the strong energy condition is violated within a finite radial interval. Thus, the exponential halo corresponds to a physically meaningful anisotropic matter distribution, but one that necessarily departs from the strong-energy-condition regime over part of the geometry. 

\noindent The halo also leaves clear imprints on observable strong-gravity signatures. The photon sphere and shadow radius increase with \(M_0\) and decrease with \(r_0\), which allows the EHT shadow measurements of M87* and Sgr A* to place approximate upper bounds on the allowed halo parameters. For scalar perturbations, the effective potential has a single barrier peak whose shape depends on the halo parameters, and the Pad\'e-resummed WKB method provides stable quasinormal frequencies across the parameter scan. In this sector, stronger or more extended halos lower both the oscillation frequency and the damping rate, producing longer-lived ringdown modes, while the standard WKB approximation becomes unreliable for strong-halo, higher-overtone configurations. 

\noindent Finally, the greybody analysis shows that the halo suppresses transmission through the effective potential barrier, especially at low and intermediate frequencies, so scalar emission becomes less efficient in the presence of a denser or more extended halo. Taken together, our results show that an exponential dark matter halo can leave measurable phenomenological signatures in black-hole geometry, curvature, shadow formation, quasinormal ringing, and greybody transmission. This makes the model a useful analytic laboratory for exploring how environmental dark matter can modify the observational and dynamical properties of black holes in astrophysical settings.

\section*{Acknowledgments}
This research was funded by the Science Committee of the Ministry of Science and Higher Education of the Republic of Kazakhstan (Grant No. AP23488743).


\bibliography{0references}
\bibliographystyle{apsrev4-1}

\end{document}